\documentclass[11pt,journal,letterpaper,onecolumn]{IEEEtran}
\usepackage{color}  % to have colored text
\usepackage[final]{graphicx} % to be able to add graphics
\usepackage{amsmath,amssymb,amscd,latexsym,dsfont,amsthm} % for mathematical 
\usepackage[skip=2pt,font=footnotesize]{caption}
\usepackage{psfrag}
\usepackage{epstopdf} % to include eps  figures.
\usepackage[pdf]{pstricks}
\usepackage{tabls}
\usepackage{cite}
\usepackage{color,comment}
\usepackage{balance}
\usepackage{float}
\usepackage{placeins}

\newcommand{\tr}[1]{\mathrm{#1}}

\newcommand{\mf}[1]{\mathsf{#1}}

\newcounter{tempEquationCounter}
\newcounter{thisEquationNumber}
{\setcounter{equation}{\value{tempEquationCounter}}% set back to equation number in floated location
\end{figure*}% end float environment
}%

\begin{document}
%%------------------
\title{Consumption Factor Optimization for Multihop Relaying over Nakagami-$m$ Fading channels}

\author{%
Itsikiantsoa Randrianantenaina, Mustapha Benjillali,~\IEEEmembership{Senior Member,~IEEE}, and~Mohamed-Slim~Alouini,~\IEEEmembership{Fellow,~IEEE}\\
\thanks{%
I. Randrianantenaina and M.-S. Alouini are with the Computer, Electrical, and Mathematical Science and Engineering (CEMSE) Division, King Abdullah University of Science and Technology (KAUST), Thuwal, Saudi Arabia. [e-mails: \{itsikiantsoa.randrianantenaina, slim.alouini\}@kaust.edu.sa].}%
\thanks{%
M. Benjillali is with the Communications Systems Department at the National Institute of Telecommunications (INPT), Rabat, Morocco. [e-mail: benjillali@ieee.org].}%
\thanks{%
This work was funded in part by King Abdullah University of Science and Technology (KAUST).
}%
}%
%%-----------------

\maketitle
\thispagestyle{empty}
   
\begin{abstract}
In this paper, the energy efficiency of multihop relaying over Nakagami-$m$ fading channels is investigated. The ``consumption factor'' is used as a metric to evaluate the energy efficiency, and it is derived in closed-form for both amplify-and-forward  and  decode-and-forward relaying. Then, based on the obtained expressions, we propose a power allocation strategy maximizing the consumption factor. In addition, two sub-optimal, low complexity, power allocation algorithms are proposed 
and analyzed, and the obtained power allocation schemes are compared, in terms of energy efficiency as well as other common performance metrics, to other power allocation schemes from the literature. Analytical and simulation results confirm the accuracy of our derivations, and assess the performance gains of the proposed approach.
\end{abstract} 
   

\newpage
%---------------------------------------------------------------------
%---------------------------------------------------------------------
\section{Introduction}
%---------------------------------------------------------------------
%---------------------------------------------------------------------

The problem of energy efficiency is one of the current biggest challenges towards green radio communications. In fact, as a result of the  introduction of new attractive technologies (new  devices, mobile streaming video, online game development, and social networking), there has been a massive increase of the number of cellular network subscribers during the last two decades. Therefore, there has been an exponential expansion of the cellular network market, followed by an increase in the number of base stations  that have dominant energy requirements.  Consequently, energy consumption by cellular networks and wireless service providers is nowadays a serious concern. In addition, from the users side, electromagnetic radiation is at its limit in many contexts, while for battery-powered devices, transmit and circuit energy consumption has to be minimized for better battery lifetime and performance.

On the other hand, the concept of multihop communications (where the source communicates with the destination via many intermediate nodes) has been revisited and adapted to mitigate wireless channel impairments and ensure broader coverage \cite{Hasna2003}. In~\cite{Li2002}, the authors have shown that, in addition to extending coverage, overcoming shadowing and reducing the transmit power, multihop communications can increase the capacity of the network at a low additional cost.

In this work,  the energy efficiency (EE) of multihop communications is analyzed using the ``consumption factor'' (CF) introduced in \cite{Murdock2014} as a metric. The performance analysis and/or optimization of EE in the context of wireless networks have been investigated in the recent literature. A performance bound analysis of the signal-to-noise ratio (SNR), the average error and the outage were presented in \cite{Karagiannidis2004} for different types of fading models such as Rayleigh, Nakagami-$m$ and Rice. The authors of \cite{Vergados2010} presented several route selection methods in multihop communications and evaluated their performance in terms of spectral efficiency (SE) and EE. The instantaneous trade-off between the total energy consumption-per-bit and the end-to-end rate under spatial reuse in wireless multihop networks was developed and analyzed in \cite{Bae2010}. For given relays positions, it was shown that the total energy consumption-per-bit is minimized by optimally selecting the end-to-end rates.  The authors of \cite{Tralli2005} investigated the basic trade-offs between energy consumption, hop distance and robustness against fading. The authors expressed the energy consumption rate as a function of the source--destination distance and derived optimal hop distances. 

Here, we first derive explicit and closed-form expressions of CF, i.e., the average number of bits transmitted per unit of power consumed by the end-to-end multihop communication system over Nakagami-$m$ fading links. We consider both non regenerative amplify-and-forward (AF) and regenerative decode-and-forward (DF) relays. The obtained CF expressions are then used to derive a CF-optimal power allocation (PA) strategy maximizing the EE for both relaying techniques. In addition, two sub-optimal, low complexity, power allocation algorithms are proposed and analyzed, and the results are compared to other power allocation strategies.

The rest of the paper is organized as follows. The system model is presented in Section~\ref{sec:System_model}. In Section~\ref{sec:CF_derivation}, we present the derivation of the CF for both AF and DF cases, and the power allocation for CF optimization is derived in Section~\ref{sec:CF_optimization}. Numerical results are presented in Section~\ref{sec:Numerical_results}. Section~\ref{sec:conclusion} concludes the paper.

%---------------------------------------------------------------------
%---------------------------------------------------------------------
\section{System Model and Notation}\label{sec:System_model}
%---------------------------------------------------------------------
%---------------------------------------------------------------------
We consider a source $\tr{R}_{0}$ and a destination $\tr{R}_{N}$ communicating through $(N-1)$ AF or DF half-duplex relays. The distance between  $\tr{R}_{0}$ and $\tr{R}_{N}$ is fixed and it is denoted by $D$.
%as shown in Fig.~\ref{fig_syst_model}. The distance between  $\tr{R}_{0}$ and $\tr{R}_{N}$ is fixed and it is denoted by $D$.
%%%%%%%%%%%%%%%%%%%%%%%%%%%%%%%%%%%%%%%%%
%\begin{figure}[h]
%\psfrag{R0}[c][c][1.4]{~\!$\tr{R}_0$}
%\psfrag{R1}[c][c][1.4]{~\!$\tr{R}_0$}
%\psfrag{R2}[c][c][1.4]{~\!$\tr{R}_1$}
%\psfrag{R3}[c][c][1.4]{~\!$\tr{R}_2$}
%\psfrag{Rn}[c][c][1.4]{~\!$\tr{R}_{N}$}
%\psfrag{N0}[c][c][1.45]{}%{~\!$\tr{N}_0$}
%\psfrag{a1}[c][c][1.45]{~\!$\alpha_1$}
%\psfrag{a2}[c][c][1.45]{~\!$\alpha_2$}
%\psfrag{an}[c][c][1.45]{~\!$\alpha_N$}
%\psfrag{d1}[c][c][1.45]{~\!$d_1$}
%\psfrag{d2}[c][c][1.45]{~\!$d_2$}
%\psfrag{dn}[c][c][1.45]{~\!$d_N$}
%\psfrag{D}[c][c][1.45]{$D$}
%\psfrag{Source}[c][c][1.45]{}%{Source}
%\psfrag{Destination}[c][c][1.45]{}%{Destination}
%\psfrag{Relay}[c][c][1.45]{}%{Relay}
%\begin{center}
%\scalebox{0.65}
%{\includegraphics{System_Model_3.eps}}
%\caption{The adopted multihop system model with $N-1$ AF relays.}
%\label{fig_syst_model}
%\par\end{center}
%\end{figure}
%%%%%%%%%%%%%%%%%%%%%%%%%%%%%%%%%%%%%%%%%%

Each node uses only the information received from its immediate predecessor. All links are considered to be independent and not necessarily identically distributed Nakagami-$m$ fading. The $i$-th hop link is of length $d_{i}$, has $\nu_{i}$ as pathloss exponent, $m_i$ as fading parameter, and $\alpha_{i}$ as instantaneous channel coefficient. Throughout the analysis, and without any loss of generality, fading parameters $m_i$ are assumed to be integer, and the noise over all channels is zero-mean additive white gaussian (AWGN) with the same variance~$N_{0}$.
 
Each node is in one of three possible states: transmission, reception, or idle mode. The power consumed by node $\tr{R}_{i}$ during a transmission phase is given by $P_{i}^\tr{t}/\varepsilon+P_{i}^\tr{ct}$, where $P_{i}^\tr{t}$ is the transmit power used by $\tr{R}_{i}$, $\varepsilon$ is the power amplifier's efficiency ($\varepsilon\!\in\ \!]0,1]$), and $P_{i}^\tr{ct}$ is the circuit power consumption (PC) in transmission mode. Similarly, we denote by $P_{i}^\tr{cr}$ and $P_{i}^\tr{ci}$ the respective PCs of $\tr{R}_{i}$ during the reception mode and the idle mode.  We denote by $\gamma_{i}=\displaystyle\frac{\left|\alpha_{i}\right|^{2}P_{i}^\tr{t}}{N_{0}(d_{i})^{\nu_{i}}}$ and $\overline{\gamma}_{i}=\mathbb{E}\left[\gamma_i\right]$ the instantaneous and average SNRs of the $i^{\tr{th}}$ hop, respectively, where $\mathbb{E}[.]$ denotes the mathematical expectation.

We adopt the following notation in the rest of the paper. $\textrm{U}\left(a, b, z\right)=\dfrac{1}{\Gamma\left(a\right)}\displaystyle\int_{0}^{\infty}\!\!\!\!\mathrm{e}^{-zt}t^{a-1}\left(1+t\right)^{b-a-1}\mathrm{e}^t~\!\tr{d}t$ denotes the confluent hypergeometric function \cite{Gradshteyn2007}. $\Gamma(v)=\displaystyle\int^{\infty}_{0}x^{v-1}\mathrm{e}^{-x} \mathrm{d}x $ and
$\gamma(v,w)=\displaystyle\int^{w}_{0}x^{v-1}\mathrm{e}^{-x}\mathrm{d}x$ denote the Gamma and lower incomplete Gamma functions, respectively.

%---------------------------------------------------------------------
%---------------------------------------------------------------------
\section{Consumption Factor Derivation}\label{sec:CF_derivation}
%---------------------------------------------------------------------
%---------------------------------------------------------------------

CF defines the energy efficiency as the maximum achievable rate (given by Shannon's capacity) per unit of power consumed to transmit (considering the transmit power itself and the circuit powers). It can be expressed in our context as
\begin{equation}\label{CF_simple}
\mf{CF}=\dfrac{B\log_{2}(1+\gamma_\tr{e2e})}{P_{\tr{tot}}},
\end{equation}
where $B$ is the total channel bandwidth and $\gamma_\tr{e2e}$ is the end-to-end SNR from $R_{0}$ to $R_{N}$,

%---------------------------------------------------------------------
\subsection{Amplify-and-Forward}
%---------------------------------------------------------------------
For AF, the end-to-end SNR is derived in \cite{Hasna2003} as
\begin{equation}\label{eq:ge2e}
\gamma_\tr{e2e}^\tr{AF}=\!\left[\prod_{i=1}^{N}\!\left(1+\frac{1}{\gamma_{i}}\right)-1\right]^{-1},
\end{equation}
and the total consumed power for an end-to-end transmission is
\begin{equation}
P_\tr{tot}^\tr{AF}=\dfrac{1}{\varepsilon}\sum_{i=0}^{N-1}P_{i}^{\tr{t}}+P_{\tr{c}}+P_{\tr{c}}^{\tr{AF}},
\end{equation}
where $P_{\tr{c}}$ represents all the circuit powers (during transmission, reception and idle modes) from $R_{0}$ to $R_{N}$, i.e., $P_{\tr{c}}=\sum_{i=0}^{N-1}P_{i}^\tr{ct} +\sum_{i=1}^{N}P_{i}^\tr{cr}+\sum_{i=0}^{N-1}P_{i}^\tr{ci}$. $P_{\tr{c}}^{\tr{AF}}$ is the additional power consumed to amplify the received signal before forwarding. Therefore, the average CF is given by
\begin{equation}\label{def_cf_bar}
\overline{\mf{CF}}^{\tr{AF}}=b_{N}\!\intop_{0}^{\infty}\!\!\ldots\!\!\intop_{0}^{\infty}\!\mf{CF}^\tr{AF}\!\left(N\right)\prod_{i=1}^{N}\!\left(\gamma_{i}^{m_{i}-1}\mathrm{e}^{-\frac{m_{i}\gamma_{i}}{\overline{\gamma}_{i}}}\right)\tr{d}\gamma_{i},
\end{equation}
where  
\begin{equation}
b_{N}=\prod\limits _{i=1}^{N}\frac{1}{\Gamma\!\left(m_{i}\right)}\!\left(\frac{m_{i}}{\overline{\gamma}_{i}}\right)^{\!m_{i}},
\end{equation}
and
\begin{equation}
\mf{CF}^\tr{AF}\!\left(N\right)=\dfrac{B\log_{2}(1+\gamma_\tr{e2e}^\tr{AF})}{P_\tr{tot}^\tr{AF}}.
\end{equation}

After a few manipulations (cf. Appendix), we get
\begin{equation}\label{eq:CF_average_AF}
\overline{\mf{CF}}^\tr{AF}=c_{N}^\tr{AF}\sum_{j=1}^{\infty}\frac{1}{j}\prod_{i=1}^{N}\frac{\Gamma\!\left(j+m_{i}\right)}{m_{i}!}\textrm{U}\!\left(j{+}m_{i}, 1{+}m_{i}, \dfrac{m_{i}}{\overline{\gamma}_{i}}\right)\!,
\end{equation}
with 
\begin{equation}
c_{N}^\tr{AF}=\dfrac{B}{\ln(2)\left[\dfrac{1}{\varepsilon}\sum\limits _{i=0}^{N-1}P_{i}^{\tr{t}}+P_{\tr{c}}+P_{\tr{c}}^{AF}\right]}.
\end{equation}
Note that the infinite sum in~\eqref{eq:CF_average_AF} converges quickly for $j\geq 10$.

%---------------------------------------------------------------------
\subsection{Decode-and-Forward}
%---------------------------------------------------------------------
For DF, the end-to-end SNR is given by
\begin{equation}
\gamma_\tr{e2e}^\tr{DF}=\displaystyle\min_{i=1 \dots N}\gamma_{i},
\end{equation}
and
\begin{equation}
P_\tr{tot}^\tr{DF}=\dfrac{1}{\varepsilon}\sum_{i=0}^{N-1}P_{i}^{\tr{t}}+P_{\tr{c}}+P_{\tr{c}}^{\tr{DF}},
\end{equation}
$P_{\tr{c}}^{\tr{DF}}$ being the additional power consumed to decode the received signal before forwarding. 

On the other hand, the cumulative distribution function (CDF) of $\gamma_\tr{e2e}^\tr{DF}$ is given by
\begin{align}
F_{\gamma_\tr{e2e}^\tr{DF}}(x)=\tr{Pr}\left(\displaystyle\min_{i=1 \dots N}\gamma_{i} \leq x \right)= 1- \tr{Pr}(\displaystyle\min_{i=1 \dots N}\gamma_{i} \geq x ) = 1- \prod_{i=1}^{N}\!\tr{Pr}\left (\gamma_{i} \geq x \right) =  1- \prod_{i=1}^{N}\left(1-F_{\gamma_{i}}(x)\right),
\end{align}
where $F_{\gamma_{i}}(.)$ is the CDF of the individual SNR of the $i$-th hop, given by
\begin{equation} \label{eq:CDF_SNRi_DF}
F_{\gamma_{i}}(x)=\frac{1}{\Gamma\!\left(m_{i}\right)} \gamma\!\left( m_{i}, \frac{m_{i}x}{\overline{\gamma}_{i}}  \right).
\end{equation}

The average CF for DF is then
\begin{equation}
\overline{\mf{CF}}^\tr{DF}=c_{N}^\tr{DF}\mathbb{E}\left[\ln\left(1+\gamma_\tr{e2e}^\tr{DF}\right)\right],
\end{equation}
where 
\begin{equation}
c_{N}^\tr{DF}=\dfrac{B}{\ln(2)\left[\dfrac{1}{\varepsilon}\sum\limits _{i=0}^{N-1}P_{i}^{\tr{t}}+P_{\tr{c}}+P_\tr{c}^\tr{DF}\right]}.
\end{equation}
Recalling that for non-negative random variables \cite{Ross2006}
\begin{equation}\label{eq:expectation_formula}
\mathbb{E}[X]=\int_0^\infty \tr{Prob}(X \ge x)\; \mathrm{d}x,
\end{equation}
we can write
\begin{equation}
\tr{Pr}\left(\ln\left(1+\gamma_\tr{e2e}^\tr{DF}\right)\geq x\right) = 1 - F_\tr{e2e}^\tr{DF}\left(\mathrm{e}^{x}-1\right).
\end{equation}
Using \eqref{eq:expectation_formula} and ~\eqref{eq:CDF_SNRi_DF}, we get
\begin{equation}\label{eq:CF_DF_average1}
\!\overline{\mf{CF}}^\tr{DF}\!=c_{N}^\tr{DF}\int_0^\infty\prod_{i=1}^{N}\left(1-\frac{\gamma\!\left(m_{i}, \dfrac{m_{i}(\mathrm{e}^{x}-1)}{\overline{\gamma}_{i}}\right)}{\Gamma\!\left(m_{i}\right)} \right)\tr{d}x.\!\!
\end{equation}
which can be computed numerically through the Gauss-Laguerre quadrature \cite{Gradshteyn2007}.

%---------------------------------------------------------------------
\subsubsection*{Special Case of Rayleigh Fading}
%---------------------------------------------------------------------
For Rayleigh fading links, $\gamma_{i}$ is exponentially distributed with parameter $1/\overline{\gamma}_{i}$. The distribution of $\gamma_\tr{e2e}^\tr{DF}$ is also exponential with parameter $\lambda_\tr{e2e}=\sum_{i=1}^{N} \frac{1}{\overline{\gamma}_{i}}$. In this case,
\begin{equation}
\overline{\mf{CF}}^\tr{DF}=c_{N}^\tr{DF}\int_0^\infty \lambda_\tr{e2e} \ln(1+x)\mathrm{e}^{-\lambda_\tr{e2e} x} \mathrm{d}x.
\end{equation}
Using \cite[(4.337-1)]{Gradshteyn2007} and $\tr{E}_{1}(x)=-\tr{Ei}(-x)$, where $\tr{Ei}(\cdot)$ is the exponential integral function, we get the expression of CF in a simple closed-form as
\begin{equation}
\overline{\mf{CF}}^\tr{DF}=c_{N}^\tr{DF}\mathrm{e}^{\lambda_\tr{e2e}}\tr{E}_{1}(\lambda_\tr{e2e}).
\end{equation}

%---------------------------------------------------------------------
%---------------------------------------------------------------------
\section{Consumption Factor Optimization}\label{sec:CF_optimization}
%---------------------------------------------------------------------
%---------------------------------------------------------------------
In this section, we derive an energy-efficient, CF-optimal, transmit power allocation strategy for the analyzed multihop relaying setup.

%---------------------------------------------------------------------
\subsection{Optimal Power Allocation}
%---------------------------------------------------------------------
Given a total power constraint $P_\tr{tot}$, the optimization problem can be formulated as follows
\begin{equation}
\begin{aligned}\label{eq:Optimization_problem}
& \underset{P_{i}^{\tr{t}}}{\tr{min}} -\overline{\mf{CF}}(P_{0}^{\tr{t}},P_{1}^{\tr{t}}, \ldots,P_{N-1}^{\tr{t}}) \\
& \tr{s.t.~~} \sum\limits _{i=1}^{N}P_{i}^{\tr{t}} \leq P_{\tr{tot}},
\end{aligned}
\end{equation}
and the Lagrangian of the problem is given by
 \begin{equation}
\mathcal{L}(P^{\tr{t}},\mu)=-\overline{\mf{CF}}(P_{0}^{\tr{t}},P_{1}^{\tr{t}}, \ldots,P_{N-1}^{\tr{t}})+\mu(\sum\limits _{i=1}^{N}P_{i}^{\tr{t}}-P_{\tr{tot}}),
\end{equation}
where $\mu$ is the Lagrange multiplier corresponding to the inequality constraint. The Karush-Kuhn-Tucker (KKT) conditions can be expressed as
\begin{equation} \label{eq:KKT}
\begin{cases}
-\dfrac{\partial \overline{\mf{CF}}}{\partial 	P_{i}^{\tr{t}}}+\mu=0, \quad i=1,\ldots,N\\
~\sum\limits _{i=1}^{N}P_{i}^{\tr{\tr{t}}} \leq P_{\tr{tot}}. 
\end{cases}
\end{equation}	
	
Note that solving the equations system in~\eqref{eq:KKT} using Newton's method is quite complex as the expressions of the first and second derivatives of $\overline{\mf{CF}}$ are not straightforward. Alternatively, in this work, we adopt the ``Automatic Differentiation'' (AD)\footnote{For MATLAB$^{{\tiny \textregistered}}$, AD was implemented by R.~D.~Neidinger in~2008 through the {\it valder} class which implements AD by operator overloading: it computes the first order derivative or multivariable gradient vectors starting with a known simple {\it valder} and propagating it through elementary functions and operators.} to compute the gradient of the objective function, then it is passed to  MATLAB's$^{{\tiny \textregistered}}$ {\it fmincon}\footnote{MATLAB's$^{{\tiny \textregistered}}$ {\it fmincon} is a powerful method for solving constrained optimization problems. However, it is not fast enough for a considerable number of hops in our context; especially for the AF case.} as one of the parameters, while using the {\it interior-point} algorithm to further accelerate the calculation time.

Note that, in addition to the increasing computation complexity (with increasing $N$), the optimization problem in~\eqref{eq:Optimization_problem} has to be solved by a central unit aware of the channel statistics of all hops, which then broadcasts the optimal transmit powers to the relaying nodes. To avoid this, we further propose a low complexity decentralized algorithm yielding close-to-optimal transmit powers.

%---------------------------------------------------------------------
\subsection{Low Complexity Suboptimal Power Allocation}\label{subsec:lcsopa}
%---------------------------------------------------------------------
The idea is to assume that each link has the same statistics of the following hop. The first node solves the optimization problem \eqref{eq:Optimization_problem}, assuming that the rest of the links have the same channel statistics as the first hop. Therefore, the optimization can be done with only one variable. The calculated optimal value is the operating transmit power of the first node. Then, the first node transmits its corresponding term in the expression of CF (calculated with the transmit power and the channel statistic) to the next node. Once the second node receives the information from the first, it formulates a new optimization problem using the information obtained from the first node and assuming that all the following hops have the same channel statistics as the second hop. The process continues until the last node.

Practically, at the $n^\tr{th}$ node $R_{n-1}$, first, the following optimization problem is solved
\begin{equation}\label{eq:Optimization_problem_at_every_node}
\begin{aligned}
& \qquad \qquad \underset{x}{\tr{max~~}} \overline{\mf{CF}}(x) \\
& \text{s.t.~~} (N-n+1)x \leq P_{\tr{tot}}-P_{\tr{tot},n-1}.
\end{aligned}
\end{equation}
where
\begin{equation}
P_{\tr{tot},n-1}=\sum\limits _{i=1}^{n-1}P_{i}^{\tr{t}}.
\end{equation}

For AF, the expression of $\overline{\mf{CF}}^\tr{AF}\!(x)$ is given by
\begin{equation}
\overline{\mf{CF}}^{\tr{AF}}(x) = c_{N}^\tr{AF}\!(x)\sum_{j=1}^{J}\frac{1}{j}\mathcal{T}_{n-1}^{\tr{AF}}(j)
 \left(\frac{\Gamma\!\left(j+m_{n}\right)}{\Gamma\!\left(m_{n}\right)}\textrm{U}\!\left(j{+}m_{n}, 1{+}m_{n}, \dfrac{m_{n}N_{0}(d_{n})^{\nu}}{\overline{\gamma}_n x}\right)\!\right)^{\! N-n+1}
\label{eq:CF_AF_x}
\end{equation}
where we consider only the first $J$ terms for the infinite sum in~\eqref{eq:CF_average_AF}, and
\begin{equation}
c_{N}^\tr{AF}\!(x)=\dfrac{B/\ln(2)}{\dfrac{1}{\varepsilon}\!\left((N-n+1)x+\sum\limits _{i=1}^{n-1}P_{i}^{\tr{t}}\right)+P_{\tr{c}}+P_{\tr{c}}^{\tr{AF}}},
\end{equation}
and
\begin{equation}
\mathcal{T}_{n-1}^{\tr{AF}}(j)=\prod_{i=1}^{n-1}\frac{\Gamma\!\left(j+m_{i}\right)}{\Gamma\!\left(m_{i}\right)}~\!\tr{U}\!\!\left(j{+}m_{i}, 1{+}m_{i}, \dfrac{m_{i}}{\overline{\gamma}_{i}}\right)\!.
\end{equation}

For DF, using a Laguerre-Gauss quadrature with $K$ terms, the expression of $\overline{\mf{CF}}^{\tr{DF}}(x)$ is given by
\begin{equation}
\overline{\mf{CF}}^\tr{DF}\!(x) \approx c_{N}^\tr{DF}\!(x)\sum_{k=1}^{K} w_{k} \mathrm{e}^{x_{k}}\mathcal{T}_{n-1}^{\tr{DF}}(k)\!\left(1-    \dfrac{\gamma\!\left(    m_{n}, \dfrac{m_{n}N_{0}(d_{n})^{\nu}(\mathrm{e}^{x_{k}}-1)}{\overline{\gamma}_n x}  \right)}{\Gamma\!\left(m_{n}\right)}     \right) ^{\! N-n+1}
\label{eq:CF_DF_x}
\end{equation}
where
\begin{equation}
c_{N}^\tr{DF}\!(x)=\dfrac{B/\ln(2)}{\dfrac{1}{\varepsilon}\!\left((N-n+1)x+\sum\limits _{i=1}^{n-1}P_{i}^{\tr{t}}\right)+P_{\tr{c}}+P_{\tr{c}}^{\tr{DF}}},
\end{equation}
and
\begin{equation}
\mathcal{T}_{n-1}^{\tr{DF}}(k)=\prod_{i=1}^{n-1}\!\!\left(1-\dfrac{\gamma\!\left(  m_{i}, \dfrac{m_{i}(\mathrm{e}^{x_{k}}-1)}{\overline{\gamma}_{i}}  \right)}{\Gamma\!\left(m_{i}\right)}     \right).
\end{equation}

After solving the optimization problem in \eqref{eq:Optimization_problem_at_every_node},  the computed optimal value of $x$ is the operating transmit power $P_{n}^{\tr{t}}$ for the $n^{\text{th}}$ node. This node transmits to the next node $P_{\tr{tot},n}=P_{\tr{tot},n-1}+ P_{n}^{\tr{t}}$ and $\mathcal{T}_{n}^{\tr{AF}}(j)$ or $\mathcal{T}_{n}^{\tr{DF}}(k)$ for $j=1,2,\ldots ,J$ or $k=1,2,\ldots ,K$. Note that, recursively, we can write
\begin{equation}
\mathcal{T}_{n}^{\tr{AF}}(j)=\mathcal{T}_{n-1}^{\tr{AF}}(j)\frac{\Gamma\!\left(j+m_{n}\right)}{m_{n}!}~\!\textrm{U}\!\!\left(j{+}m_{n}, 1{+}m_{n}, \dfrac{m_{n}}{\overline{\gamma}_{n}}\right),
\end{equation}
and
\begin{equation}
\mathcal{T}_{n}^{\tr{DF}}(k)=\mathcal{T}_{n-1}^{\tr{DF}}(k)\!\!\left(1-    \dfrac{\gamma\!\left(    m_{n}, \dfrac{m_{n}(\mathrm{e}^{x_{k}}-1)}{\overline{\gamma}_{n}}  \right)}{\Gamma\!\left(m_{n}\right)}     \right).
\end{equation}

%--------------------------------------------------------------------------------------
%--------------------------------------------------------------------------------------
\section{Performance Results and Comparisons}\label{sec:Numerical_results}
%--------------------------------------------------------------------------------------
%--------------------------------------------------------------------------------------

%----------------------------------------------------
\subsection{Comparison Framework}\label{sec:compset}
%----------------------------------------------------
For performance comparison purposes, and in order to complete the analysis, we first define another power allocation strategy based on the optimization of the end-to-end capacity.
The end-to-end capacity of the analyzed system was already implicitly derived in Sec.~\ref{sec:CF_derivation} as it is the numerator in the CF expression. Explicitly, for AF, it can be expressed as
\begin{equation}
\mf{C}^\tr{AF}=\frac{B}{\ln(2)}\sum_{j=1}^{\infty}\frac{1}{j}\prod_{i=1}^{N}\frac{\Gamma\!\left(j+m_{i}\right)}{m_{i}!}\textrm{U}\!\left(j{+}m_{i}, 1{+}m_{i}, \dfrac{m_{i}}{\overline{\gamma}_{i}}\right)\!,
\end{equation}
and, for DF, with analogy to~\eqref{eq:CF_DF_average1} and using a $K$-order Gauss-Laguerre quadrature, it can be written as
\begin{equation}
\mf{C}^\tr{DF}\approx \frac{B}{\ln(2)}\sum_{k=1}^{K} w_{k} \mathrm{e}^{x_{k}}\prod_{i=1}^{N}\left(1-\frac{\gamma\!\left(m_{i}, \dfrac{m_{i}(\mathrm{e}^{x_{k}}-1)}{\overline{\gamma}_{i}}  \right)}{\Gamma\!\left(m_{i}\right)}\right)\!,
\end{equation}
where $x_{k}$ and $w_{k}$ are the sample points and the weight factors of the Laguerre polynomial \cite{Gradshteyn2007}.

Based on these expressions, we can derive a capacity-optimal power allocation scheme as follows
\begin{equation}\label{eq:Optimization_capacity}
\begin{aligned}
& \underset{P_{i}^\tr{t}}{\tr{min}} -\mf{C}^\tr{AF/DF}(P_{0}^\tr{t},P_{1}^\tr{t}, \ldots,P_{N-1}^\tr{t}) \\
& \tr{s.t.~~} \sum\limits _{i=1}^{N}P_{i}^\tr{t} \leq P_\tr{tot}^\tr{AF/DF}.
\end{aligned}
\end{equation}

We thus discuss the performance of the proposed CF-based approach in~\eqref{eq:Optimization_problem}, i.e., the CF optimizing PA (denoted as CFoPA), and we compare it to the results obtained with other PAs strategies, namely: {\it i)} a first low complexity CF-based suboptimal uniform PA (CFsoUPA) where all nodes are constrained to transmit with the same power, computed to maximize the CF \cite{Randrianantenaina2014}, {\it ii)} a second low complexity CF-based suboptimal PA (CFsoPA) presented in~\ref{subsec:lcsopa}, {\it iii)} a capacity-optimal PA (CoPA) defined in~\eqref{eq:Optimization_capacity}, and finally {\it iv)} a uniform PA (UPA) where the total power is just equally divided between transmitting nodes, i.e., $P^{\tr{t}}_{i}=P_{\tr{tot}}/N$, as a reference case.

%-------------------------------------
\subsection{Numerical Results}
%-------------------------------------
We consider that the total distance between the source and the destination is normalized to unity, and that the relays are positioned uniformly between the source and the destination. The system bandwidth is also normalized. For numerical results, the pathloss exponent is $\nu{=}4$, the power amplifier's efficiency is $\epsilon{=}0.35$ and, in order to alleviate the calculations without any loss of generality, the total circuit power for AF relays is $P_{c}\!+\!P_{c}^{\tr{AF}}\!=\!0.3N$, and $P_{\tr{c}}\!+\!P_{\tr{c}}^{\tr{DF}}\!=\!0.4N$ for DF relays; unless it is clearly specified otherwise.

%---------------------------------------------------------------------
\subsubsection{Effect of The Number of Hops} 
\begin{figure}[ht]
\captionsetup{justification=centering}
\psfrag{m=2}[cl][cl][1.2]{~\!$m=2$}
\psfrag{m=1}[cl][cl][1.2]{~\!$m=1$}
\psfrag{PAOCF}[cl][cl][1]{\!CFoPA}
\psfrag{PASOCF}[cl][cl][1]{\!CFsoUPA}
\psfrag{CFsoPA}[cl][cl][1]{\!CFsoPA}
\psfrag{UPA}[cl][cl][1]{\!UPA}
\psfrag{PAOC}[cl][cl][1]{\!CoPA}
\psfrag{Nhops}[cl][cl][1.2]{$N$}
\psfrag{CF-AF}[cl][cl][1.2]{$\overline{\mf{CF}}^\tr{AF}$}
\psfrag{CF-DF}[cl][cl][1.2]{$\overline{\mf{CF}}^\tr{DF}$}
\begin{center}
\begin{tabular}{l}
\hskip -0.2cm
\scalebox{0.6}{\includegraphics{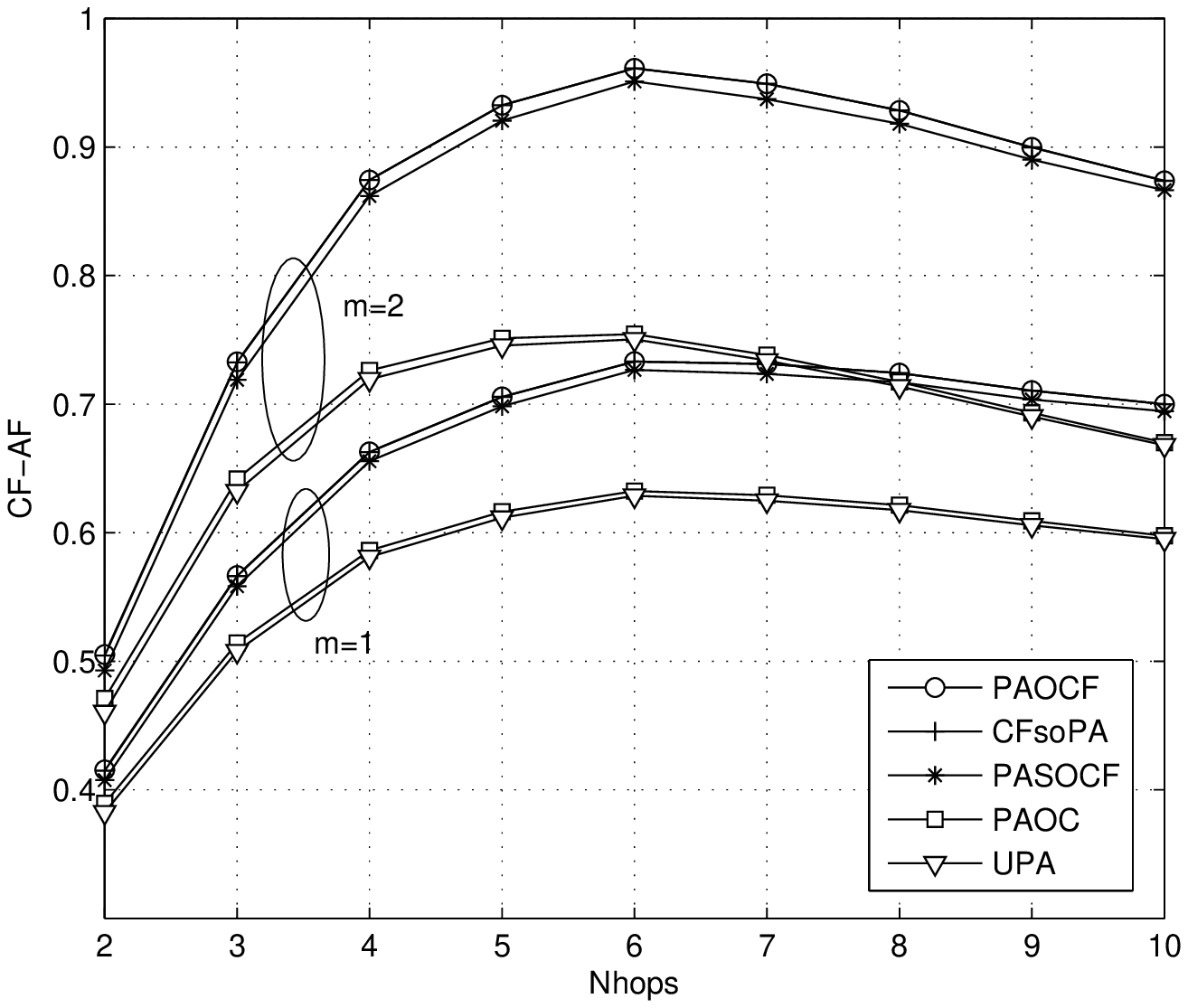}} 
\scalebox{0.6}{\includegraphics{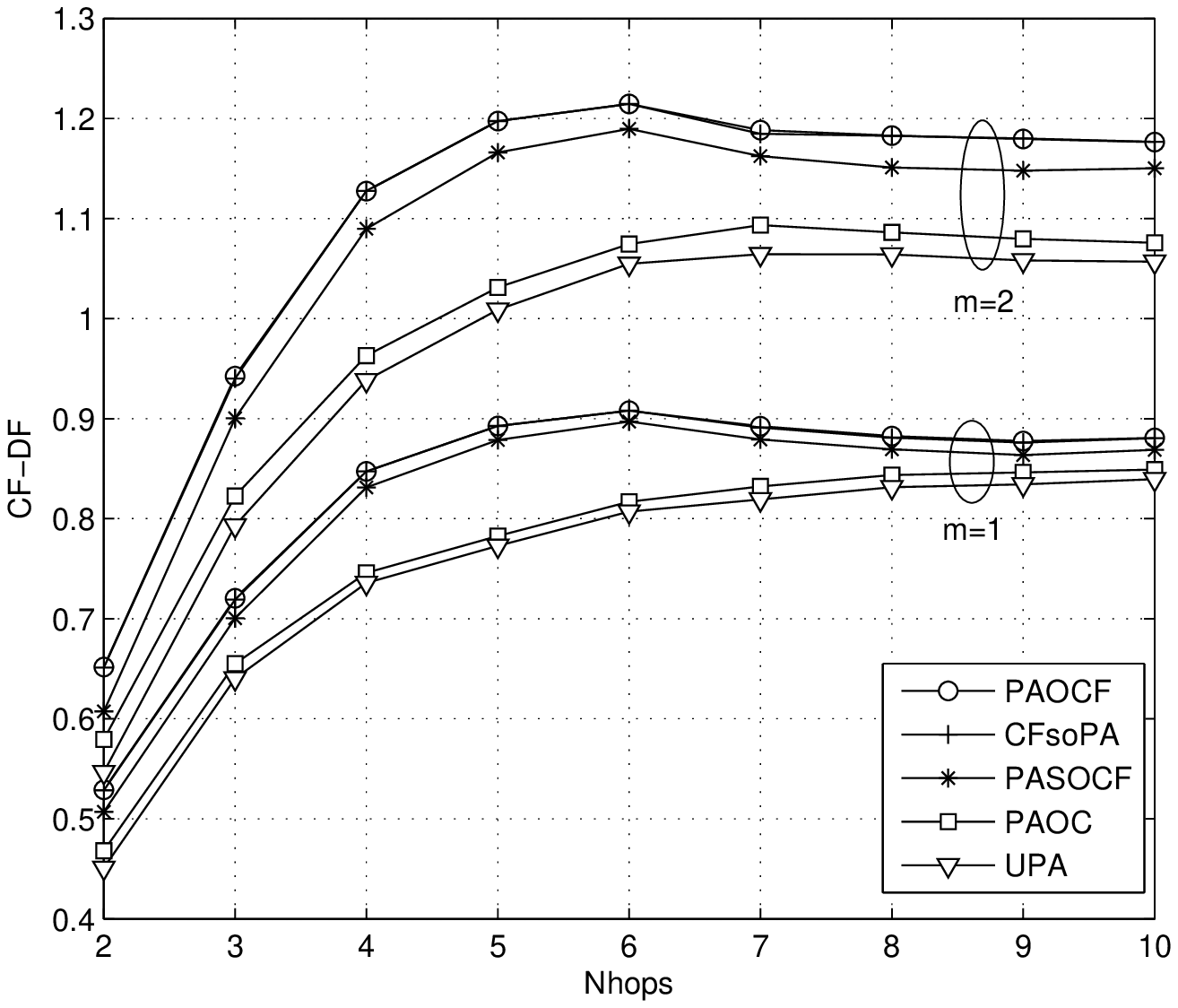}}
\end{tabular}
\end{center}
\vskip -0.4cm
\caption{Variation of $\overline{\mf{CF}}$ with the number of hops for both AF and DF and different PA schemes. $P_\tr{tot}^\tr{AF}=P_\tr{tot}^\tr{DF}=0$~\!dB.}
\label{fig:CF_Nhops}
\end{figure}

%---------------------------------------------------------------------
Fig.~\ref{fig:CF_Nhops} shows the variation of $\overline{\mf{CF}}$ with the number of hops for different PA techniques and different values of $m$. It can be seen that, for both AF and DF, and for all PA techniques, operating with a non optimal number of hops can considerably affect the performance in terms of CF, with losses up to about 50\%. We note that the figure also confirms the performance of the simple CFsoPA proposed in~\ref{subsec:lcsopa} which is practically the same as the optimal CFoPA, and slightly outperforming the sub-optimal CFsoUPA. 
%%%%%%%%%%%%%%%%%%%%%%%%%%%%%%%%%%%%%%%%%
%%%%%%%%%%%%%%%%%%%%%%%%%%%%%%%%%%%%%%%%%

%On the other hand, Fig.~\ref{fig:Effect_of_Pc} shows a comparison of the variation of $\overline{\mf{CF}}$ for AF and DF depicting the effect of the circuit power on  $\overline{\mf{CF}}$. For the sake of a  fair comparison, we consider $P_\tr{c}^{i}\!=(P_{\tr{c}}\!+\!P_{\tr{c}}^{\tr{AF}})/N$ for AF, and for DF, $P_{\tr{c}}^{i}\!=(P_{\tr{c}}\!+\!P_{\tr{c}}^{\tr{DF}})/N$. It can be seen that DF is more energy efficient than AF. The figure shows also that, as the circuit power per node is increased, $\overline{\mf{CF}}$ decreases in addition to a lower CF-optimal number of hops.
%%%%%%%%%%%%%%%%%%%%%%%%%%%%%%%%%%%%%%%%%%
%\begin{figure}[ht]	
%\psfrag{CF}[cl][cl][1.4]{$\overline{\mf{CF}}$}
%\psfrag{Nhops}[cl][cl][1.4]{$N$}
%\psfrag{AF}[cl][cl][1]{AF}
%\psfrag{DF}[cl][cl][1]{DF}
%\psfrag{Pc=0.4}[cl][cl][1]{$P_{\tr{c}}^{i}\!=\!0.4$}
%\psfrag{Pc=0.5}[cl][cl][1]{$P_{\tr{c}}^{i}\!=\!0.5$}
%\psfrag{Pc=0.6}[cl][cl][1]{$P_{\tr{c}}^{i}\!=\!0.6$}
%\psfrag{Pc=0.3}[cl][cl][1]{$P_{\tr{c}}^{i}\!=\!0.3$}
%\begin{center}
%\scalebox{0.55}{\hskip -0.8cm\includegraphics{Effect_of_Pc.eps}}
%\caption{Comparison of $\overline{\mf{CF}}^\tr{AF}$ and $\overline{\mf{CF}}^\tr{DF}$ for different circuit powers. $m=2$ and $P_\tr{tot}^\tr{AF}=P_\tr{tot}^\tr{DF}=0$~\!dB. }
%\label{fig:Effect_of_Pc}
%\par\end{center}
%\end{figure}
%%%%%%%%%%%%%%%%%%%%%%%%%%%%%%%%%%%%%%%%%%
%---------------------------------------------------------------------
\subsubsection{Effect of The Power Budget}
\begin{figure}[h!]
\captionsetup{justification=centering}
\psfrag{m=2}[cl][cl][1.2]{\!\!\!\!$m=2$}
\psfrag{m=1}[cl][cl][1.2]{~\!$m=1$}
\psfrag{PAOCF}[cl][cl][1]{CFoPA}
\psfrag{PASOCF}[cl][cl][1]{CFsoUPA}
\psfrag{CFsoPA}[cl][cl][1]{CFsoPA}
\psfrag{UPA}[cl][cl][1]{UPA}
\psfrag{PAOC}[cl][cl][1]{CoPA}
\psfrag{CF-AF}[cl][cl][1.2]{$\overline{\mf{CF}}^{\tr{AF}}$}
\psfrag{CF-DF}[cl][cl][1.2]{$\overline{\mf{CF}}^{\tr{DF}}$}
\psfrag{Ptot-AF}[cl][cl][1.2]{$P_\tr{tot}^\tr{AF}$~\![dB]}
\psfrag{Ptot-DF}[cl][cl][1.2]{$P_\tr{tot}^\tr{DF}$~\![dB]}
\psfrag{Ptot}[cl][cl][1.2]{$P_\tr{tot}^\tr{\textcolor{red}{AF}}$~\![dB]}
\psfrag{CF}[cl][cl][1.2]{$\overline{\mf{CF}}^\tr{\textcolor{red}{DF}}$}
\begin{center}
\begin{tabular}{l}
\hskip -0.4cm
\scalebox{0.6}{\includegraphics{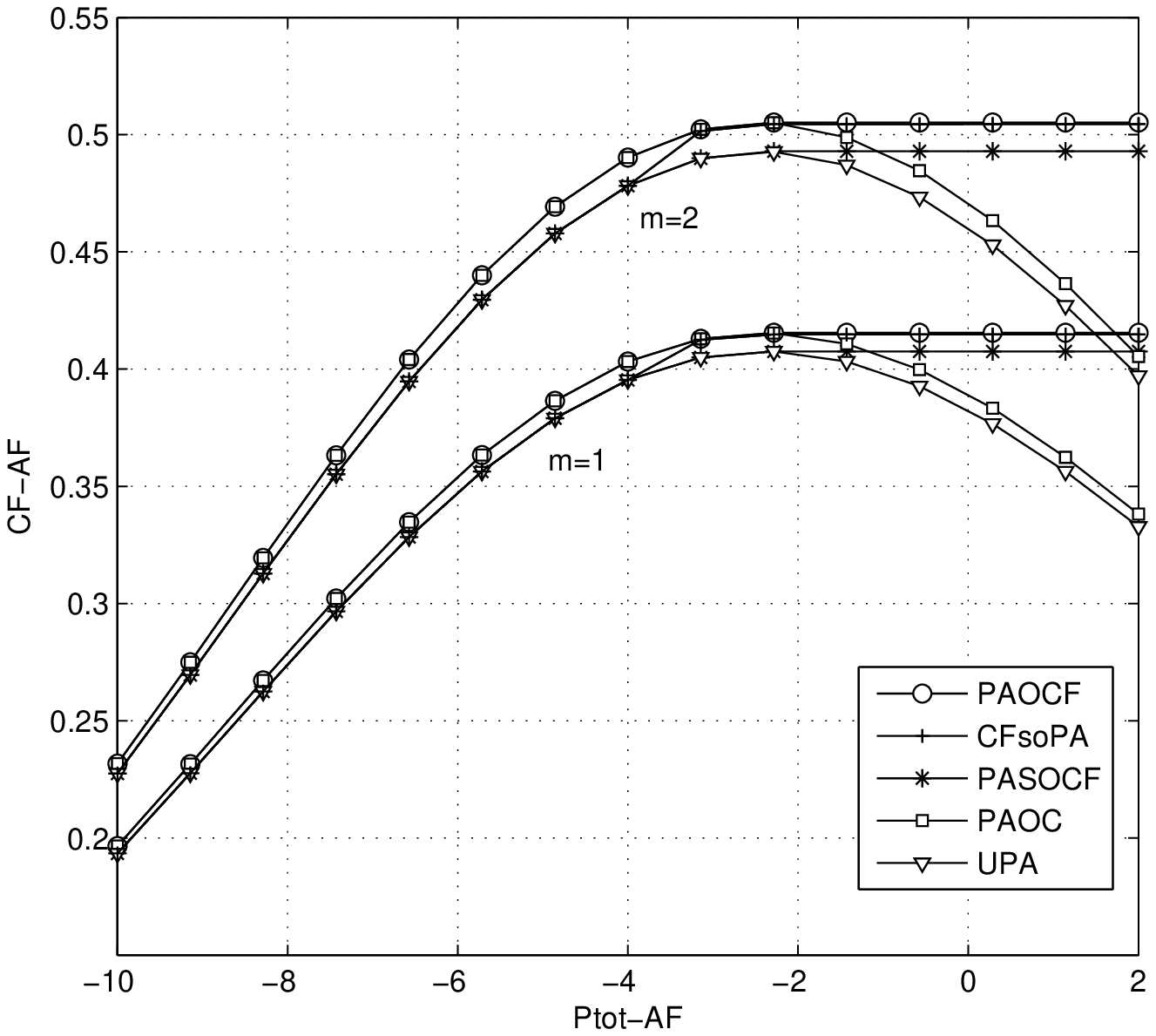}}
\scalebox{0.59}{\includegraphics{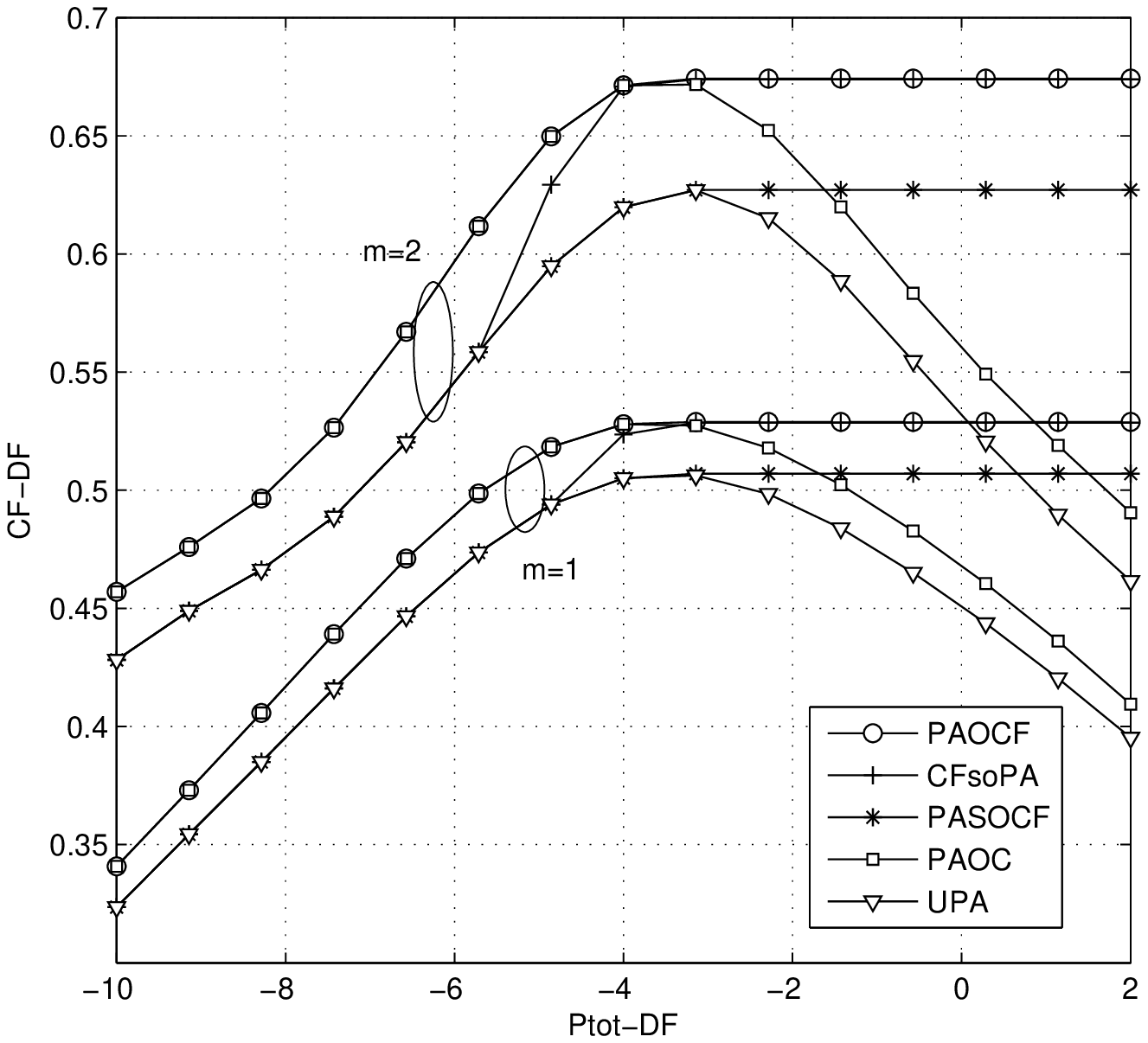}}
\end{tabular}
\end{center}
\vskip -0.4cm
\caption{Variation of $\overline{\mf{CF}}$ according to the total transmit power budget, for different PA schemes ($N=2$).}
\label{fig:CF_Ptot}
\end{figure}

%---------------------------------------------------------------------
Fig.~\ref{fig:CF_Ptot} shows $\overline{\mf{CF}}$ as a function of the total power budget for different PA techniques. It can be observed that CFoPA and CoPA, and similarly UPA and CFsoUPA, yield similar results for high to average power constraints. The performance of non CF-optimizing methods then decreases for high power budgets. This can be explained by the fact that CoPA is operating with the maximum available transmit power with no constraints on the transmit rate, the constraint in (\ref{eq:Optimization_capacity}) is hence always satisfied at the boundary of the feasible region, i.e., $\sum_{i=1}^{N}P_{i}^\tr{t} = P_{\tr{tot}}$.
%%%%%%%%%%%%%%%%%%%%%%%%%%%%%%%%%%%%%%%%%
%%%%%%%%%%%%%%%%%%%%%%%%%%%%%%%%%%%%%%%%%

\begin{figure}[ht]
\captionsetup{justification=centering}
\psfrag{m=2}[cl][cl][1]{~\!{$m\!=\!2$}}
\psfrag{m=1}[cl][cl][1]{~\!{$m\!=\!1$}}
\psfrag{CF-AF}[cl][cl][1.2]{$\overline{\mf{CF}}^\tr{AF}$}
\psfrag{CF-DF}[cl][cl][1.2]{$\overline{\mf{CF}}^\tr{DF}$}
\psfrag{capacity-AF}[cl][cl][1.2]{$\mf{C}^\tr{AF}$}
\psfrag{capacity-DF}[cl][cl][1.2]{$\mf{C}^\tr{DF}$}
\begin{center}
\begin{tabular}{l}
\hskip -0.4cm
\scalebox{0.6}{\includegraphics{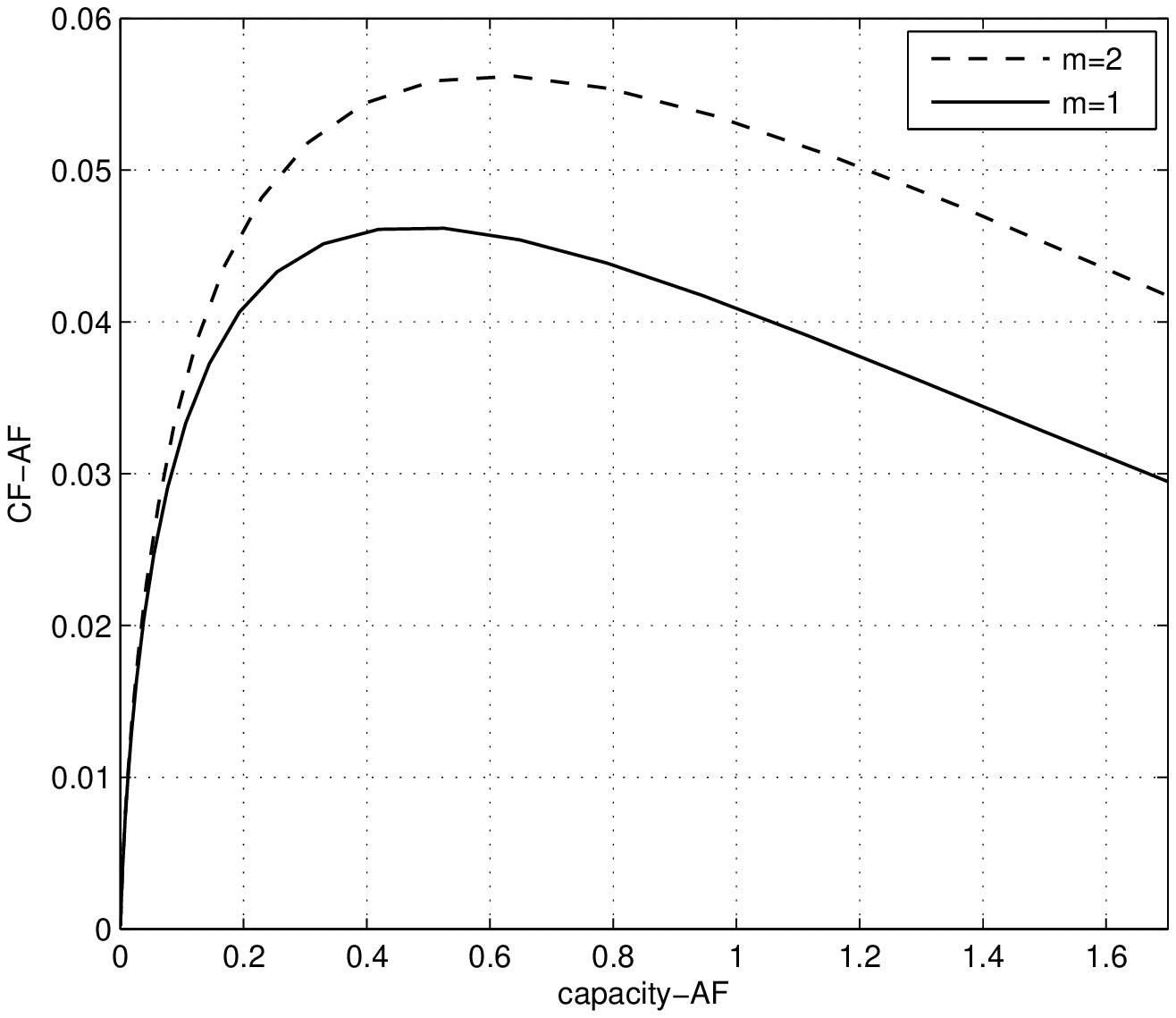}}
\scalebox{0.6}{\includegraphics{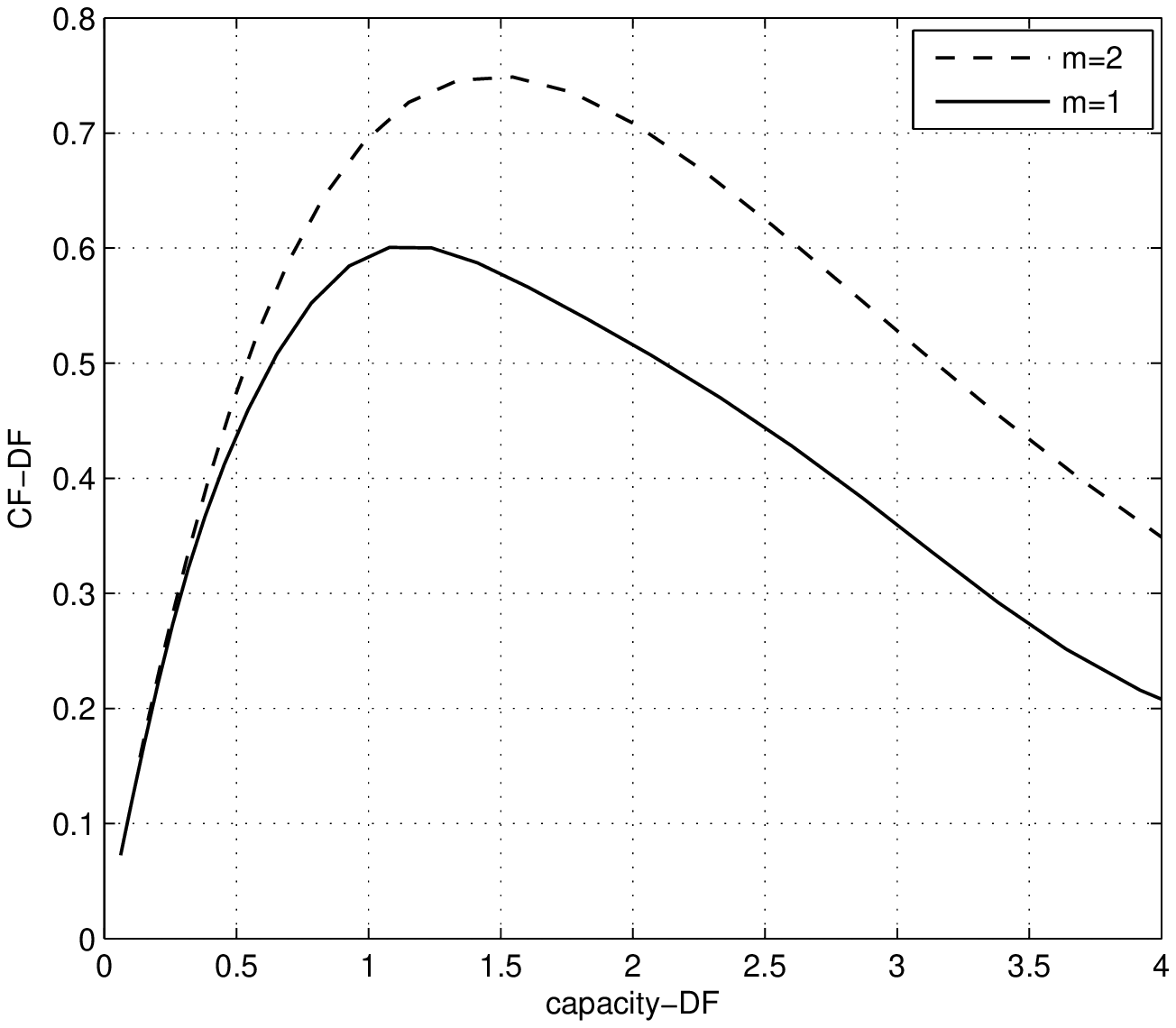}}
\end{tabular}
\end{center}
\vskip -0.4cm
\caption{Tradeoff between the $C$(Capacity) and the $\overline{\mf{CF}}$, $\overline{\gamma}_1=\overline{\gamma}_2=1$, $N=2$ and $P_{\tr{tot}}=0$dB .}
\label{fig:Tradeoff}
\end{figure}

For relatively low power budgets, the power term in the denominator of the expression of CF is limited even at the maximum transmit power allocation. The CF optimization in~\eqref{eq:Optimization_problem} is therefore a maximization of the capacity (numerator in the expression of CF), using the entire transmit power budget. Above a given power budget threshold, referred to as the {\it critical power budget}, all increase in the transmit power, while increasing the capacity, is decreasing the CF ratio. The CF-based optimal PA is thus to allocate a fixed transmit power beyond that critical point, outperforming the capacity-optimal strategy which keeps transmitting at the increasingly maximum available transmit power. This capacity-CF tradeoff is presented in Fig.~\ref{fig:Tradeoff} for both AF and DF.
%%%%%%%%%%%%%%%%%%%%%%%%%%%%%%%%%%%%%%%%%
%%%%%%%%%%%%%%%%%%%%%%%%%%%%%%%%%%%%%%%%%

%---------------------------------------------------------------------
\subsubsection{Impact of Dissimilar Link Conditions}
\begin{figure}[ht]	
\captionsetup{justification=centering}
\psfrag{DF}[cl][cl][1]{DF}
\psfrag{CF}[cl][cl][1.2]{$\overline{\mf{CF}}$}
\psfrag{AF}[cl][cl][1]{AF}
\psfrag{PAOCF}[cl][cl][1]{\!CFoPA}
\psfrag{CFsoPA}[cl][cl][1]{\!CFsoPA}
\psfrag{PASOCF}[cl][cl][1]{\!CFsoUPA}
\psfrag{Difference}[cl][cl][1.2]{$\Delta_{\overline{\gamma}}=\overline{\gamma}_1-\overline{\gamma}_2$}
\begin{center}
\scalebox{0.65}
{\hskip -0.8cm\includegraphics{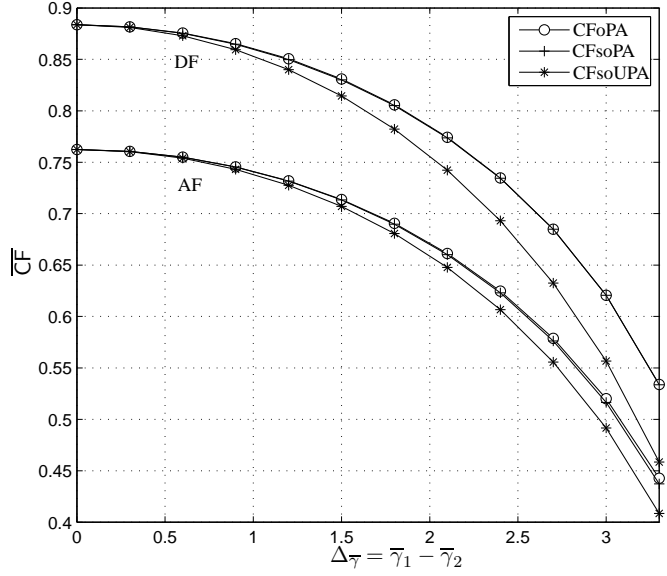}}
\caption{Effect of dissimilar fading on $\overline{\mf{CF}}$ with $P_\tr{tot}^\tr{AF}=P_\tr{tot}^\tr{DF}=0$~\!dB for $N=2$.}
\label{fig:Channel_amplitude_effect}
\par\end{center}
\end{figure}

%---------------------------------------------------------------------
From the above discussion, it is obvious that the difference between CFoPA and CFsoUPA, and similarly the difference between CoPA and UPA, will not be notable for identically distributed hops. In order to show the impact of the link conditions on the performance of the sub-optimal power strategies, we consider the particular case of a dual-hop relaying.  Fig.~\ref{fig:Channel_amplitude_effect} shows a comparison between CFoPA and CFsoUPA in terms of $\overline{\mf{CF}}$ for both AF and DF when the first and the second hops experience different fading conditions, e.g., $\overline{\gamma}_1\neq \overline{\gamma}_2$. We can observe that the loss in performance of the sub-optimal CFsoUPA increases when the difference $\Delta_{\alpha}=\overline{\gamma}_1- \overline{\gamma}_2$ is increasing (we assume that $\overline{\gamma}_1> \overline{\gamma}_2$ without loss of generality).
%%%%%%%%%%%%%%%%%%%%%%%%%%%%%%%%%%%%%%%%%
%%%%%%%%%%%%%%%%%%%%%%%%%%%%%%%%%%%%%%%%%

%---------------------------------------------------------------------
%---------------------------------------------------------------------
\section{Conclusion} \label{sec:conclusion}
%---------------------------------------------------------------------
%---------------------------------------------------------------------
In this paper, closed-form expressions of the energy consumption factor of multihop relaying for both AF and DF relays were derived. A power allocation technique maximizing this CF metric was then proposed, and its performance (in terms of EE and capacity) was compared to other PA schemes from the literature. 

The analysis of the numerical results shows an interesting tradeoff between CF and the end-to-end capacity, and the existence of a critical operating transmit power budget. In addition, it was shown that the number of hops should be defined carefully taking into consideration both end-to-end power constraints and individual circuit powers of the relaying nodes. Finally, an in all investigated scenarios, it was observed that DF presents a relative advantage compared to AF from the energy efficiency point of view.

Among the many possible extensions of the actual work, the analysis of the critical transmit power budget, and the derivation of the optimal number of hops over other types of fading and/or in other spectrum sharing contexts would be of high interest to complete the investigation of multihop relaying schemes.

%---------------------------------------------------------------------
%---------------------------------------------------------------------
\appendix[Derivation of $\overline{\mf{CF}}^\tr{AF}$ in \eqref{eq:CF_average_AF}]
%---------------------------------------------------------------------
%---------------------------------------------------------------------
From the definition in~\eqref{def_cf_bar}, we have
\begin{equation}
\overline{\mf{CF}}^\tr{AF}= b_{\!N}\!\intop_{0}^{\infty}\!...\!\intop_{0}^{\infty}\mf{CF}^\tr{AF}\!\left(N\right)\prod_{i=1}^{N}\!\left(\gamma_{i}^{m_{i}-1}\mathrm{e}^{-m_{i}\gamma_{i}/\overline{\gamma}_{i}}\tr{d}\gamma_{i}\right),
\end{equation}
where
\begin{align}
\mf{CF}\!\left(N\right)&=\frac{B}{P_\tr{tot}^\tr{AF}\ln(2)} \ln\!\!\left(\! 1+\!\left[\prod_{i=1}^{N}\!\left(\dfrac{\gamma_{i}+1}{\gamma_{i}}\right)-1\right]^{-1}\right)=-\frac{B}{P_\tr{tot}^\tr{AF}\ln(2)}\ln\!\left(\dfrac{\prod\limits _{i=1}^{N}\!\left(\gamma_{i}+1\right)-\prod\limits _{i=1}^{N}\gamma_{i}}{\prod\limits _{i=1}^{N}\!\left(\gamma_{i}+1\right)}\right)=-\frac{B\ln\!\left(1-f_{N}\right)}{P_\tr{tot}^\tr{AF}\ln(2)},
\end{align}
and  $f_{N}=\prod\limits _{i=1}^{N}\gamma_{i}/\prod\limits _{i=1}^{N}\!\left(1+\gamma_{i}\right)$. Using the Taylor series expansion
$\ln\!\left(1-x\right)=-\sum\limits _{j=1}^{\infty}\dfrac{x^{j}}{j}, \ \tr{~for~}\!\left|x\right|<1$, 
and noting that
\begin{align}
\intop_{0}^{+\infty}\!\!&\left(\!\frac{\gamma_{i}}{1+\gamma_{i}}\!\right)^{\!j}\!\gamma_{i}^{m_{i}-1}\mathrm{e}^{-\frac{m_{i}\gamma_{i}}{\overline{\gamma}_{i}}}d\gamma_{i}=\!\left(\!\dfrac{m_{i}}{\overline{\gamma}_{i}}\!\right)^{\!\!-m_{i}}\Gamma\!\left(j+m_{i}\right)\textrm{U}\!\!\left(j+m_{i}, 1+m_{i}, \dfrac{m_{i}}{\overline{\gamma}_{i}}\right),
\end{align}
we can finally write 
\begin{align}
&\overline{\mf{CF}}_{\tr{AF}}
\!=\!\frac{B}{P_\tr{tot}^\tr{AF}\ln(2)}\sum_{j=1}^{\infty}\frac{1}{j}\prod_{i=1}^{N}\frac{\Gamma\!\left(j+m_{i}\right)}{\Gamma\!\left(m_{i}\right)}\textrm{U}\!\!\left(\!j{+}m_{i}, 1{+}m_{i}, \dfrac{m_{i}}{\overline{\gamma}_{i}}\!\right)\!.
\end{align}
	
%---------------------------------------------------------------------
%---------------------------------------------------------------------
\bibliography{Biblio_IEEE_TVT}
\bibliographystyle{IEEEtran}
%---------------------------------------------------------------------
%---------------------------------------------------------------------

\end{document}